\documentclass{article}
\usepackage{ltwol2e}
\def\eqref#1{(\ref{#1})}

\usepackage{psfig}
\usepackage{epsfig}

\def\Journal#1#2#3#4{{#1} {\bf #2}, #3 (#4)}

\def\NPB{{\em Nucl. Phys.} B}
\def\PLB{{\em Phys. Lett.}  B}
\def\PRL{\em Phys. Rev. Lett.}
\def\PRD{{\em Phys. Rev.} D}
\def\ZPC{{\em Z. Phys.} C}
\def\JHEP{\em JHEP}
\def\IJMPA{{\em Int. J. Mod. Phys.} A}
\def\epj{{\em Eur. Phys. J. } C}
\def\ibid#1#2#3{{\em ibid.} {\bf #1}, #3 (#2)}

\def\z0{\rm Z^0}
\def\p{\rm p}

\newcommand{\epem}{\rm e^+\rm e^-}
\newcommand{\amz}{\as(M^2_{\rm Z})}
\def\eps{\epsilon}
\def\as{\alpha_s}
\def\al{\alpha}
\def\cb{\beta}
\def\lrang#1{\left\langle#1\right\rangle}
\def\cO#1{{\cal{O}}\left(#1\right)}
\def\cM{{\cal{M}}}
\def\cA{{\cal{A}}}
\def\cP{{\cal{P}}}
\def\half{\textstyle \frac12}

\def\cF{{\cal{F}}}
\def\cR{{\cal{R}}}
\def\cFd{\dot{\cal{F}}}
\def\muI{\mu_I}
\def\ae{\alpha_{\mbox{\scriptsize eff}}}
\def\bk{{\bf k}}
\def\PT{\mbox{\scriptsize{PT}}}
\def\NP{\mbox{\scriptsize{NP}}}
\def\UV{{\mbox{\scriptsize UV}}}
\def\IR{{\mbox{\scriptsize{IR}}}}
\def\MSbar{\overline{\mbox{\scriptsize MS}}}
\def\LQCD{\Lambda_{\mbox{\scriptsize QCD}}}

\begin{document}

\title{PERTURBATIVE QCD THEORY
(INCLUDES OUR KNOWLEDGE OF $\as$)}

\author{YU. L. DOKSHITZER}

\address{INFN, sezione di Milano, via G. Celoria 16, 
 20133 Milan, Italy \\E-mail: yuri.dokshitzer@mi.infn.it \\
and \\
 St. Petersburg Nuclear Physics Institute, 188350 Gatchina,
  St. Petersburg, Russia}

\twocolumn[\maketitle
\abstracts{
The problem of what we know, think we know, and think about
the QCD coupling $\as$ is discussed. }]

\section{The Ground gives rise to Measurements}
QCD has a split personality, ``Perturbative QCD'' and
``Non-perturbative aspects of QCD''  being routinely separated by
the organisers of HEP conferences. And so they are in our minds\footnote{ 
A reader willing to refresh his/her awareness of deep puzzles of the
game is kindly advised to consult the Proceedings of last year's HEP EPS
conference.\cite{EPS97}}.  
The microscopic dynamics of quarks and gluons is the QCD battleground; 
understanding the spectrum and interactions of hadrons 
is its ultimate objective.
 
The objective of this talk is less ambitious.
My aim is to make you aware (if not convinced) of  a possibility
of a root that starts off with the QCD Lagrangian, employs a
good old Dyson-Feynman field-theoretical staff of quark and gluon
Green functions and may eventually lead to an understanding of colour 
confinement. 
To embark on such a quest one should believe in legitimacy of using the
language of {\bf quarks} and {\bf gluons} down to small momentum scales, 
which implies understanding and describing the physics of 
confinement in terms of the standard QFT machinery, that is,
essentially, {\bf perturbatively}.\cite{Gribov}

Is this programme {\em crazy enough}\/ to have a chance to be correct?  
It seems {\em it is}.

\subsection{pQCD: a sketchy health report}
We shall start from a brief biased display of recent theoretical
advances on the perturbative frontier.
\begin{enumerate}
\item
Mass effects in heavy quark production 
cross sections are now available at the NLO level.\cite{NOhq}
\item A bunch of new state-of-the-art (2-loop + all-log-resummed) pQCD
  predictions have been derived to treat jet cross
  sections,\cite{jcs} hadroproduction of heavy quarks~\cite{hqhp} and
  prompt photons,\cite{gammahp} secondary heavy quark
  pairs,\cite{hqsea} the $C$-parameter~\cite{Cparam} and
  broadening~\cite{Bdist} distributions in $\epem$ annihilation.
  Techniques are being developed for addressing {\em
    next-to-next-to-leading}\/ order issues.\cite{NNLO} 
  
\item   
  Serious technical progress has been achieved in describing the
  {H}igh {E}nergy {R}egime of scattering cross sections of two small
  QCD objects, the BFKL Heron (regretfully known as the ``Hard
  Pomeron'').\cite{BFKL} This object should be responsible for a steep
  energy growth of production cross sections of

$\left\{
\mbox{
\begin{minipage}{5in}
HERA {forward jets} with {$p_t\simeq Q$}, \\
widely separated in rapidity Tevatron jets, \\
{$\gamma\gamma\to J/\psi+J/\psi$}, and alike.
\end{minipage}
}\right.$

\noindent
The origin of the large NLO correction to the BFKL evolution 
kernel~\cite{BFKLkern}
is under scrutiny: 
how much of it is due to ``kinematical'' effects 
in the evolution,\cite{BFKLkinem} 
running coupling,\cite{BFKLrunning} 
{\em angular ordering}\/ effects in space-like evolution.\cite{BFKLao} 
A physically motivated resummation of subleading kinematical and 
collinear effects 
seems to greatly improve the convergence of the \PT\ 
analysis.\cite{BFKLkinem,BFKLcollinear}
 
\item 
A unification of deep inelastic and diffractive phenomena is underway.
It employs the notion of non-forward (``off-diagonal'',
``off-forward'') 
[double] parton distributions which, on one hand, are related with various
hadron form factors~\cite{npd} 
and, on the other hand, can be accessed in 
hard interactions such as deeply virtual Compton
scattering~\cite{ji,dvcs} or hard diffractive electroproduction of
(vector) mesons.\cite{hdevm} Of special interest are 
parton-helicity-sensitive (``magnetic'') non-forward 
distributions~\cite{magnetic} that do not contribute to the usual
inclusive DIS cross sections (structure functions) but 
participate, e.g.,  in determining the contribution of the
quark orbital angular momentum to the proton spin,\cite{ji} 
see also~\cite{jirad} and references therein.

\item 
Gluon radiation induced by propagation of a colour charge through a QCD
medium (QGP, nuclear matter) attracts increasing attention.\cite{LPM}

\item 
An ideology and machinery for probing
 {\em non-perturbative}\/
 (\NP) 
effects 
with {\em perturbative}\/ 
 (\PT) 
tools is being developed. 
\end{enumerate}

\begin{table*}[t]
\caption{World summary of measurements of $\as$.
Abbreviations:
DIS = deep inelastic scattering; GLS-SR = Gross-Llewellyn-Smith sum rules;
Bj-SR = Bjorken sum rules;
(N)NLO = (next-to)next-to-leading order perturbation theory;
LGT = lattice gauge theory;
resum. = resummed next-to-leading order.}
\begin{center}
\begin{tabular}{|r l|c|l|l|c c|c|}
   \hline
 & &  Q & & &  \multicolumn{2}{c|}
{$\Delta \amz $} &  \\ 
 & Process & [GeV] & $\alpha_s(Q)$ &
  $ \amz$ & exp. & theor. & Theory \\
\hline \hline \normalsize
 & & & & & & & \\[-3mm]
& DIS [pol. strct. fctn.] & 0.7 - 8 & & $0.120\ ^{+\ 0.010}
  _{-\ 0.008}$ & $^{+0.004}_{-0.005}$ & $^{+0.009}_{-0.006}$ & NLO \\
& DIS [Bj-SR] & 1.58
  & $0.375\ ^{+\ 0.062}_{-\ 0.081}$ & $0.121\ ^{+\ 0.005}_{-\ 0.009}$ & 
  -- & -- & NNLO \\
$\bullet$ & DIS [GLS-SR] & 1.73
  & $0.295\ ^{+\ 0.092}_{-\ 0.073}$ & $0.114^{+0.010}_{-0.012}$ & 
  $^{+0.005}_{-0.006}$ & $^{+0.009}_{-0.010}$ & NNLO \\
$\bullet$& $\tau$-decays 
  & 1.78 & $0.339 \pm 0.021$ & $0.121 \pm 0.003$
  & 0.001 &  0.003 & NNLO \\
 & DIS [$\nu$; ${\rm F_2\ and\ F_3}$]  & 5.0
  & $0.215 \pm 0.016$
   & $0.119\pm 0.005$   &
    $ 0.002 $ & $ 0.004$ & NLO \\
& DIS [$\mu$; ${\rm F_2}$]
     & 7.1 & $0.180 \pm 0.014$ & $0.113 \pm 0.005$ & $ 0.003$ &
     $ 0.004$ & NLO \\
& DIS [HERA; ${\rm F_2}$]
     & 2 -- 10 &  & $0.120 \pm 0.010$ & $ 0.005$ &
     $ 0.009$ & NLO \\
$\bullet$ & DIS [HERA; jets]
     & 10 -- 100 &  & $0.118 \pm 0.009$ & $ 0.003$ &
     $ 0.008$ & NLO \\
& DIS [HERA; ev.shps.]
     & 7 -- 100 &  & $0.118\ ^{+\ 0.007}_{-\ 0.006}$ & $ 0.001$ &
     $^{+0.007}_{-0.006}$ & NLO \\
& ${\rm Q\overline{Q}}$ states
     & 4.1 & $0.223 \pm 0.009$ & $0.117 \pm 0.003 $ & 0.000 & 0.003
     & LGT \\
$\bullet$ & $\Upsilon$ decays
     & 4.13 & $0.220 \pm 0.027$ & $0.119\ \pm 0.008
     $ & 0.001 & $0.008$ & NLO \\
$\bullet$ &  $\epem$ [$\sigma_{\rm had}$] 
     & 10.52 & $0.20 \pm 0.06 $ & $0.130\ ^{+\ 0.021\ }_{-\ 0.029\ }$
     & $\ ^{+\ 0.021\ }_{-\ 0.029\ }$ & -- & NNLO \\
 & $\epem$ [ev. shapes]  & 22.0 & $0.161\ ^{+\ 0.016}_{-\ 0.011}$ &
   $0.124\ ^{+\ 0.009}_{-\ 0.006}$ &  0.005 & $^{+0.008}_{-0.003}$
   & resum \\
& $\epem$ [$\sigma_{\rm had}$]  & 34.0 &
 $0.146\ ^{+\ 0.031}_{-\ 0.026}$ &
   $0.123\ ^{+\ 0.021}_{-\ 0.019}$ & $^{+\ 0.021}_{-\ 0.019}
   $ & -- & NLO \\
$\bullet$ & $\epem$ [ev. shapes]  & 35.0 & $ 0.145\ ^{+\ 0.012}_{-\ 0.007}$ &
   $0.123\ ^{+\ 0.008}_{-\ 0.006}$ &  0.002 & $^{+0.008}_{-0.005}$
   & resum \\
$\bullet$ & $\epem$ [ev. shapes]  & 44.0 & $ 0.139\ ^{+\ 0.010}_{-\ 0.007}$ &
   $0.123\ ^{+\ 0.008}_{-\ 0.006}$ & 0.003 & $^{+0.007}_{-0.005}$
   & resum \\
& $\epem$ [ev. shapes]  & 58.0 & $0.132\pm 0.008$ &
   $0.123 \pm 0.007$ & 0.003 & 0.007 & resum \\
& $\p\bar{\p} \rightarrow {\rm b\bar{b}X}$
    & 20.0 & $0.145\ ^{+\ 0.018\ }_{-\ 0.019\ }$ & $0.113 \pm 0.011$ 
    & $^{+\ 0.007}_{-\ 0.006}$ & $^{+\ 0.008}_{-\ 0.009}$ & NLO \\
& ${\rm p\bar{p},\ pp \rightarrow \gamma X}$  & 24.2 & $0.137
 \ ^{+\ 0.017}_{-\ 0.014}$ &
  $0.111\ ^{+\ 0.012\ }_{-\ 0.008\ }$ & 0.006 &
  $^{+\ 0.010}_{-\ 0.005}$ & NLO \\
& ${\sigma (\rm p\bar{p} \rightarrow\  jets)}$  & 30 -- 500 &  &
  $0.121\pm 0.009$ & 0.001 & 0.009 & NLO \\
$\bullet$ & $\epem$ [$\Gamma (\z0 \rightarrow {\rm had.})$]
    & 91.2 & $0.122\pm 0.005$ & 
$0.122\pm 0.005$ &
   $ 0.004$ & $0.003$ & NNLO \\
& $\epem$ [ev. shapes] &
    91.2 & $0.122 \pm 0.006$ & $0.122 \pm 0.006$ & $ 0.001$ & $
0.006$ & resum \\
 & $\epem$ [ev. shapes]  & 133.0 & $0.111\pm 0.008$ &
   $0.117 \pm 0.008$ & 0.004 & 0.007 & resum \\
 & $\epem$ [ev. shapes]  & 161.0 & $0.105\pm 0.007$ &
   $0.114 \pm 0.008$ & 0.004 & 0.007 & resum \\
 & $\epem$ [ev. shapes]  & 172.0 & $0.102\pm 0.007$ &
   $0.111 \pm 0.008$ & 0.004 & 0.007 & resum \\
 $\bullet$ & $\epem$ [ev. shapes]  & 183.0 & $0.109\pm 0.005$ &
   $0.121 \pm 0.006$ & 0.002 & 0.006 & resum \\
 $\bullet$ & $\epem$ [ev. shapes]  & 189.0 & $0.109\pm 0.006$ &
   $0.122 \pm 0.007$ & 0.003 & 0.006 & resum \\
\hline
\end{tabular}
\end{center}
\end{table*}

\noindent
The latter topic does not actually belong to the list because its main
objective lies beyond what is conventionally considered to be the
perturbative domain.

\subsection{Running QCD coupling -- 1998}
A precise measurement of the coupling constant and verification of
asymptotic freedom remains a primary target for experimental QCD
studies. Not that chasing the third $\amz$ digit would answer many (if
any) a serious problem.  Still Table~1 which summarises $\as$
measurements~\cite{Siggi} is very important for the theory. We should
treat it with utmost respect as the major test of consistency of the
tools that we employ to address various aspects of interactions involving
hadrons.  The results which appeared or were updated since summer
1997~\cite{SB97} (marked with a ``$\bullet$'') include $\as$ from
\begin{enumerate}
\item the GLS sum rule, based on new CCFR $\nu$-$N$ 
      scattering data,\cite{CCFR} 
\item detailed and precise high-statistic ALEPH~\cite{tauALEPH} 
    and OPAL~\cite{tauOPAL} studies of vector and axial-vector
    channels of hadronic $\tau$-decays,   
\item H1 differential (2+1) jet rates at HERA,\cite{H1jets} 
\item $\Upsilon$-decays,\cite{Upsilon}
\item a CLEO determination from $E_{cm}$=10.52 GeV total hadronic 
      cross section measurement,\cite{CLEO}
\item reanalysis of JADE $E_{cm}$=35 and 44 GeV $\epem$ annihilation
  data to include an
  additional jet-shape variable --- the $C$-parameter,\cite{JADEalpha}
\item the most recent LEP update of the ratio of hadronic to leptonic
  $Z^0$ decay widths,\cite{LEPratio}
\item and finally, event shapes measured at the highest LEP energies,  
$E_{cm}$=183 and 189 GeV.\cite{LEP2shapes}
\end{enumerate}

\begin{figure}[t]
\begin{center}
\epsfig{figure=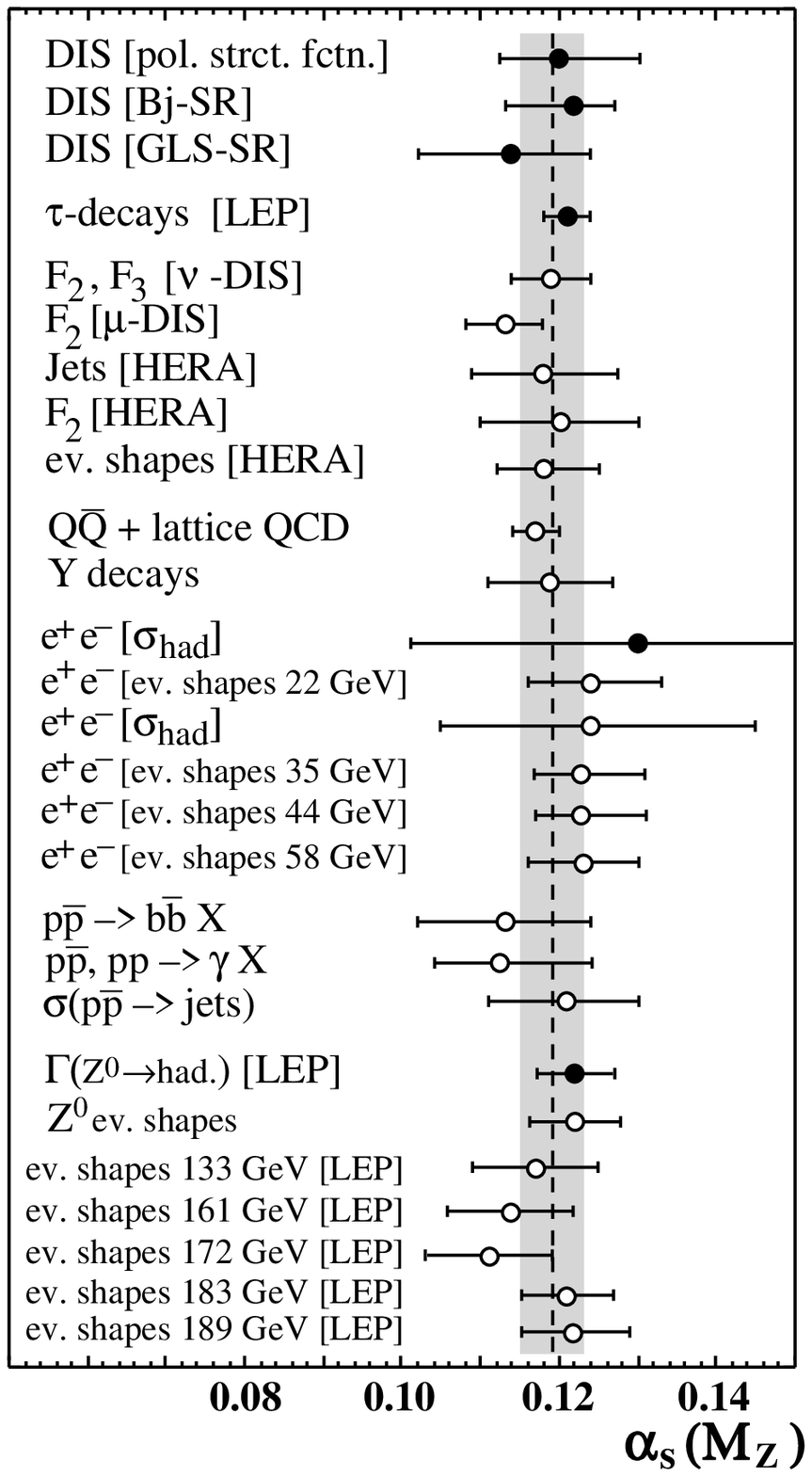,width=3in,height=3.5in}
\end{center}
\caption{Compilation of  {\protect $\amz$} measurements.
{\protect \cite{Siggi}}} 
\end{figure}

\begin{figure}[t]
\begin{center}
\epsfig{figure=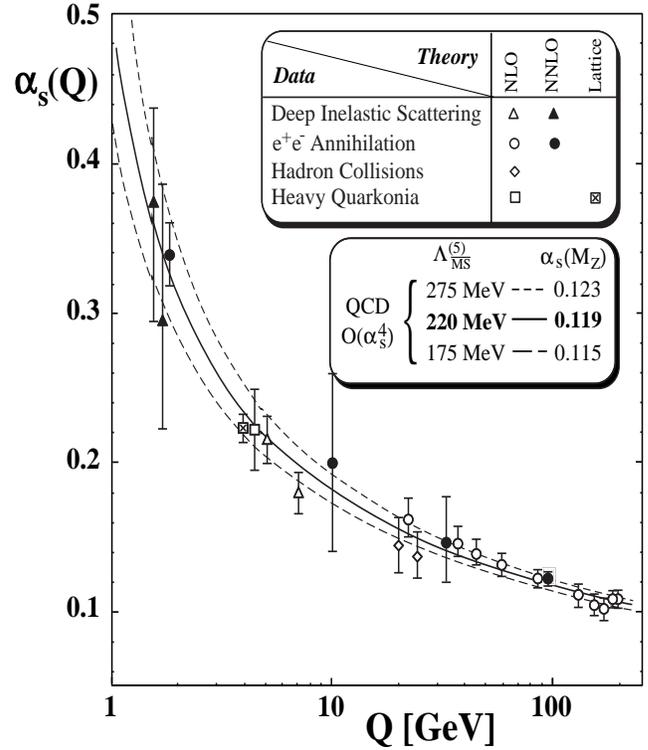,width=3.3in,height=3.5in}
\end{center}
\caption{Running QCD coupling--1998.
{\protect\cite{Siggi}}} 
\end{figure}

\noindent
The results on $\amz$
from~\cite{JADEalpha,LEPratio,LEP2shapes} are still
preliminary.  Table~1 and Fig.~1 display $\amz$ results evolved
using the 4-loop $\MSbar$ 
$\cb$-function with 3-loop matching~\cite{4loop} at
quark pole masses $M_b\!=$4.7~GeV and $M_c=$1.5~GeV.
Fig.~2 collects $\as$ values at the proper scales of individual experiments
and spectacularly demonstrates asymptotic freedom.

It is important to stress that ``all meaningful subsamples of results
provide similar average values, and there is no significant systematic
shift between any of those subsamples''.  Remarkedly, the ``only
$\epem$'' and ``only DIS'' subsamples peacefully coexist now, yielding
$\amz=$0.1210$\pm$0.0049 and 
$\amz=$0.1175$\pm$0.0061 correspondingly.
And so do ``only
$Q\le$10~GeV'' 
(0.1179$\pm$0.0043) and ``only $Q\ge$30~GeV 
(0.1208$\pm$0.0058)
measurements.

The world average value of the $\MSbar$ coupling $\amz$ 
is finally quoted to be~\cite{Siggi} 
$$
\amz = 0.119 \pm 0.004\>.
$$
This result is based on 18 (of 27) individual measurement from the
Table~1 with the errors $\Delta\amz\le 0.008$.  The overall uncertainty is
derived by the ``optimised correlation'' method.\cite{Schmelling}

\subsection{$\as$ fresh from Lattice}

A lattice analysis of the strong coupling was recently carried out by
the group~\cite{Orsay} working on a QUADRICS QH1 at Orsay.  Calculations
were performed on $16^4$ and $24^4$ lattices with a bare lattice
coupling constant $\beta\!=\!6$ (corresponding to an inverse lattice
spacing $a^{-1} \simeq 1.9$~GeV).  Calculations repeated for $\beta\!=\!6.2$
($a^{-1} \simeq 2.7$~GeV) on a $24^4$ lattice (which roughly embodies
the same physical volume as $16^4$ for $\beta\!=\!6.0$) yield consistent
results. 
The primary objective of the study was to measure the coupling   
(Landau gauge, MOM subtraction scheme) ``as defined in the
text-books''. Namely, 
\begin{enumerate}
\item
  to measure 3- and 2-point gluon correlations, the Green
  functions $G_3(p_1,p_2,-p_1-p_2)$ and $G_2(p)=Z(p^2)/p^2$,
\item 
  truncate $G_3$ to extract the vertex function \\
  $\Gamma_3(p_1,p_2,p_3)=
  G_3(p_1,p_2,p_3)\prod_{i=1}^3 G_2^{-1}(p_i^2)$ 
\item 
and then construct the coupling at a symmetric Euclidean point $k^2\!>\!0$,
$$
 \as(k^2) 
= \left. \Gamma_3(p_1,p_2,p_3)\right|_{p_1^2=p_2^2=p_3^2=-k^2}\cdot 
 Z^{3/2}(-k^2)\,.
$$
\end{enumerate}
Fig.~3 shows the result. The adjacent Fig.~4 displays the behaviour of 
a similar {\em asymmetric}\/ correlator in which one of the three gluon
momenta is set to zero. 

\begin{figure}[t]
\begin{center}
\epsfig{figure=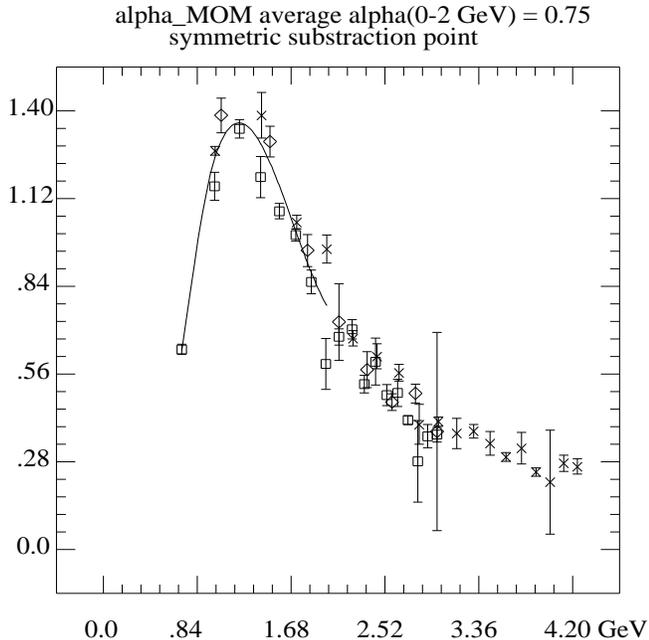,width=2.8in,height=3.5in}
\end{center}
\caption{Running coupling on the lattice.{\protect\cite{Orsay}}} 
\end{figure}

\begin{figure}[t]
\begin{center}
\epsfig{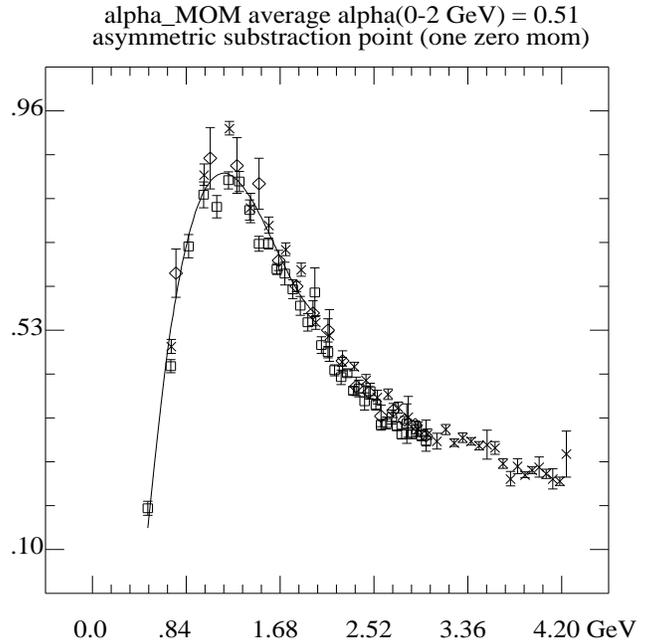}
\end{center}
\caption{``Asymmetric'' lattice coupling.{\protect\cite{Orsay}}} 
\end{figure}

The fact that the lattice coupling shows the tendency to {\em
  decrease}\/ at small momenta should not surprise us.  Regardless of
which particular kind of confinement we observe on the lattice, the
correlators of coloured fields had better vanish at the origin.  The
gluon Green function cannot have a pole at $p^2=0$, nor any similar
singularity strong enough to propagate colour flux at large
distances, $j(R)\cdot R^2=\cO{1}$.
Therefore, having chosen to truncate $G_3$ with the text-book
perturbative gluon propagators, we are bound to see $Z(p^2)$ trying to
kill the massless gluon pole.

It does not matter for our discussion how realistic these plots are. 
In particular, quarks were not included in the game. Even if they
had been, there is always a question of how to incorporate light
fermions with Compton wave lengths comparable to (exceeding) the
volume of the lattice world. This is an especially troubling
question if light quarks do indeed play a crucial r\^ole in {\em the}\/
confinement of the world we live and experiment in.\cite{Gribov}

\section{Measurements give rise to Assessments}

\subsection{The origin and status of $\as(k^2)$}

\begin{flushright}
\begin{minipage}{2.8in}
{\em  Therefore, those who are not thoroughly aware of the disadvantages
  in the use of arms cannot be thoroughly aware of the advantages
  in the use of arms.}\cite{AW}  
\end{minipage}
\end{flushright}

\noindent
We are used to the notion of a running QCD coupling.
What is its origin and status? 
Formally, $\as$ is a parameter of the perturbative expansion.  From this
point of view, its choice is, to a large extent, a matter of free
will: it depends on how smart we want to be in organising the \PT-series.
It starts to make more sense, and becomes natural, within a programme
of trading a formal expansion parameter for a {\em smarter}\/ object,
a momentum-dependent $\as(k^2)$, which embodies some specific all-order
radiative correction effects and is supposed to truly represent
the interaction strength at a given momentum scale.

A bunch of questions immediately comes to mind:  
\begin{itemize}
\item
what high-order effects to embed? 
\item 
what does ``{\em truly represent}\/'' mean?
\item
what is the characteristic momentum scale that the running coupling depends on?
\end{itemize}
Renormalisability of the theory 
is known to be responsible for the running of the
coupling. The Renormalisation Group tells us:  
``scale the whole World, and the answer remains the same provided 
$\al$ has been changed accordingly''. 
Stretching the World is not at our disposal, however. We may scale
up/down {\em external momenta}\/  but not, say, hadron masses. 
Therefore the running $\as$ rightfully appears in the realm of 
Hard Interactions characterised by a large momentum transfer, 
$Q^2\gg m^2$, and in the description of observables insensitive to 
finite InfraRed parameters (particle masses $m$).

The RG dictates how the renormalised coupling ({\em constant}\/)  
$\as(\mu_R^2)$ changes with the renormalisation scale $\mu_R$. 
It is casually said that 
in small-distance amplitudes characterised by a single large Euclidean
virtual momentum scale $Q\gg m$, the formal dependence on the renormalisation
point, $\as(\mu_R^2)$, can be traded for $\as(Q^2)$
thus giving rise to the coupling ({\em function\/}) running 
with the physical momentum. 

A couple of cautious comments are due before we go further.  Firstly,
trading $\as(\mu_R^2)$ for $\as(Q^2)$ implies that $\mu_R/Q$ is the
only momentum ratio that enters.  This is true for renormalised
off-shell amplitudes.  These are however unphysical and consequently
gauge-dependent objects; hence, the problem of peeling off
gauge-dependent effects from the running coupling.  Dealing with
``physical'' gauge-independent on-shell amplitudes has a different
problem.  They are infrared sensitive and so have nasty $\as\ln Q/m$
(collinear) and $\as\ln Q/0$ (soft) high-order corrections. Therefore,
one needs to treat collinear- and infrared-safe cross sections instead
of $n$-point amplitudes.

Secondly, the way the running coupling emerges in Minkowskian
observables is rather tricky. ``On-shell'' gluons are produced even
from small distances $1/k\ll 1$~fm with a coupling $\as(0^2)$, in a
manner of speaking, and not with $\as(k^2)$ (in a direct analogy with
real photons from $Z$-peak being radiated with $\al_{em}(0)\simeq
1/137$ rather than $\al_{em}(M_Z)\simeq 1/128$). What is instead
determined by $\as(k^2)$ is an intensity of inclusive emission of a
gluon sub-jet with the total transverse momentum $k$ with respect to
the radiating quark.

At the two-loop level the $\as(\mu_R^2)$-dependence proves to be {\em
  universal}, i.e.\  
independent of the choice of the \PT-expansion parameter. This
universality is inherited from the basic property of {\em ultraviolet
  renormalisability}\/ of the theory.  This part of the momentum
variation of $\as$ is determined by the first two terms in the
$\cb$-function,
$$
 \frac{d}{d\ln\mu_R} \left(\frac{\as(\mu_R^2)}{2\pi}\right)^{-1} = 
\cb(\as(\mu_R^2))\>, 
$$
namely,
$$
\cb(\al) \>=\> \cb_0 + \cb_1\frac{\al}{2\pi}  + \ldots
$$
The further expansion terms $\cb_{n}$ with $n\ge2$ remain 
{\em scheme}- (and even {\em gauge})-dependent; in other words, arbitrary.
Therefore, the large-momentum behaviour of the running coupling $\as(Q^2)$
cannot be uniquely fixed beyond two loops. 
The reason for that is pretty simple: only the first two loops are truly 
dominated by the \UV\ 
region, that is by small-distance physics. 

There aren't many \UV-dominated QCD diagrams.  Logarithmically
divergent radiative corrections to gluon and quark wave functions
(self energies) and to gluon-gluon (quark-gluon) interaction vertices
are involved in the renormalisation of the coupling constant $\as$.

Consider for example a quark loop in the gluon self-energy.
The one-loop radiative corrections contain the standard integral
$$
\int \frac{d^4q}{q^4}\propto \ln\Lambda_{\UV} 
\>\>  =\>\> \infty
\quad {\Longrightarrow} \quad \cb_0\>.
$$
Hiding infinity under the carpet produces $\cb_0$, the first coefficient 
in the \PT\ $\cb$-function expansion. 
In the next step we supply our quark loop with an additional internal
gluon. Now we have two independent loop-momenta to integrate over,
$q_1$ and $q_2$. Integration regions $q_{1(2)}\ll q_{2(1)}\ll \Lambda_\UV$
could have produced $(\ln\Lambda_\UV)^2$ contributions. 
These get suppressed by renormalising the {\em internal}\/ propagators and
vertices at the one-loop level, the result being 
a single-logarithmic integral determined by the region 
$q_1\sim q_2\ll \Lambda_\UV$,  
\begin{equation}
  \label{eq:oneloop}
\int \frac{d^4q}{q^4}\as(\mu_R^2)\>\>\propto\>\> 
\as(\mu_R^2) \ln\Lambda_{\UV}\>\>=\>\> \infty 
\quad {\Longrightarrow} \quad \cb_1\>.  
\end{equation}
This is how the usual story goes, order by order in perturbation theory. 
We can do better, however, by taking into consideration that the coupling in
\eqref{eq:oneloop} runs with the internal momentum.  This means
reorganising the \PT\ 
series so as to incorporate  into the two-loop diagram
the higher order effects that result in substituting
$\as(\mu_R^2)\to\as(q^2)$. 
By doing so we obtain a contribution which is still \UV-divergent, 
though modified by the logarithmic decrease of the coupling at large momenta,
$$
\int \frac{d^4q}{q^4}\as(q^2)\propto 
\ln\ln\Lambda_{\UV}\>\> =\>\> \infty 
\quad {\Longrightarrow} \quad \cb_1\>.
$$
Renormalising it out gives rise to  $\cb_1$. Starting from the third
loop the situation however changes drastically: the \UV-region is no longer
dominant, and we get
$$
 \int \frac{d^4q}{q^4}\left(\as(q^2)\right)^2\>\>=\>\> \mbox{finite}
\quad{\Longrightarrow} \quad \mbox{\begin{minipage}{1.3in}
$\cb_{n\ge2}$  depend on the 
{\bf infrared} physics!
  \end{minipage}}
$$
The notorious ``nobody is perfect'' applies to the above
consideration as well.  Strictly speaking, it is not known how to
systematically refurbish, in all orders, the \PT-series in terms of
the running $\as$. So, this argument can be looked upon as yet another
example of a statement ``correct but unproven'',\cite{Euler} see
however.\cite{Watson} Still, the message it sends is clear: starting
from the $\as^3$ (next-to-next-leading) level, a purely perturbative
treatment may become intrinsically ambiguous because of
an interconnection between small and large distances.  In particular,
there may be no way of unambiguously defining the QCD coupling $\as$,
beyond the two loops, without solving the Theory in the infrared.

\subsection{Defining $\as$ beyond two loops}
Bearing in mind this warning one can still attempt to define {\em some}\/ 
\PT\ 
expansion parameter $\as$ beyond two loops.
To this end we may try different options.

{\bf 1}. 
As long as higher terms in the $\cb$-function are arbitrary anyway, why not  
simply set $\cb_{n\ge2}\equiv0$. This is the so-called 't Hooft scheme which
is perfectly fine but for one thing: the Landau singularity 
--- the fake infrared problem --- obviously limits the use of $\as$  
so defined, to sufficiently large momentum scales.

{\bf 2}. 
One can design some sort of  ``optimisation principle'' 
in order to fix $\cb_2$. The leading idea
here is to minimise, one way 
or another, the effect of a (typically unknown, with few exceptions) 
three-loop \PT\ 
correction to a given observable    
(Minimal Sensitivity, Fastest Apparent Convergence). 
Guided by sheer pragmatism (not necessarily a curse word) 
this approach often results in fixing $\cb_2$ such as to force $\as(k^2)$ 
to develop a ``spurious infrared fixed point''. 

A (curse) word {\em spurious}\/ appears here because the 3-loop
$\cb$-function and the corresponding coupling that emerge from
such an analysis are observable-dependent. 
Linked to this is a well-grounded criticism based on non-transitivity
between the couplings defined on the base of different
observables. 
Moreover, it is legitimate to ask ourselves, whether it makes sense 
to optimise \PT\ 
series which are known to be senseless, factorially
divergent, being contaminated by {\bf infrared renormalons}. 

Given all these reservations, it is still interesting to notice 
that  such a ``spuriously finite'' coupling 
comes out close to the characteristic magnitude of 
the interaction strength in the infrared domain, the ``couplant''
$\lrang{\frac{\as}{\pi}}\simeq 0.2$, which we shall discuss below.  

{\bf 3}. 
We may try to link $\as$ with some physical quantity in a 
``most natural'' way. 
To this family belong  ``effective charges'',\cite{GrunbergEC}
specific, physically motivated, 
perturbative definitions  of $\as$ such as, e.g., the BLM,\cite{BLM} 
CMW (MC),\cite{CMW} or Uraltsev~\cite{Uraltsev}
schemes fitted to serve gluon exchange potential, relativistic gluon
radiation and non-relativistic heavy quark physics, respectively. 
The words ``most natural'' don't deceive us. 
Whether we like it or not, there is no direct link between $\as$ and
observables, the latter being {\em hadronic}\/ observables, in a
confining theory.   

{\bf 4}. 
It is perfectly allowable to play with the $\cb$-function 
(in particular, by adding to it non-analytic pieces) 
so as to make $\as(k^2)$ freeze (or even vanish) in the origin. 
Everything is allowed as long as we do not spoil the large-momentum
behaviour of the running $\as(k^2)$ at the $1/\ln k$ and $(1/\ln k)^2$
level.  

Rightfully rejecting the infrared ``Landau singularity'' 
in the coupling as being unphysical,  
or blaming infrared renormalons for spoiling \PT\ 
expansions, or discussing possible links between $\as$ and hadronic
observables we stumble upon the very same problem: what is a {\em true
  measure}\/ of interaction between {\em inexistent}\/ objects?  By
hook or by crook, the problem is that of defining the theory in the
infrared: the confinement problem.

\subsection{Does $\as$ at low momentum scales make sense?}
Possible responses are
\begin{itemize}
\item{\bf Conservative:} No way. {\em It cannot be, because this can't
    be, never.}\cite{Prutkoff}   
  As long as there are no quarks and gluons in the physical spectrum of
  the theory, we cannot talk about the QCD interaction strength at
  distances above which colour-bearers get (mysteriously) confined.
  As a conservative position it is as impeccable as it is infertile.
\item{\bf Liberal:} 
Let's try.  Defining an {\em infrared-finite}\/ coupling $\as$ 
may serve as a tool for probing universal
features of colour confinement.\cite{MS}  
({\em He who takes no risks drinks no champaign.})
\item{\bf Revolutionary:}
We must.
Dyson-Feynman's is the only reliable Field-Theoretical language in our
disposal. We have no other way but to employ quark and gluon 
Green functions down to $k\!=\!0$ and search for an unusual,
confinement, solution.
\end{itemize}

\paragraph{Practically,} God
has generously supplied us with {\em light quarks}\/ capable of
delivering {\em early screening}, preventing colour fields from going
haywire.  As shown by V.N. Gribov,\cite{Gribov} to achieve {\em
  super-critical binding}\/ of light quarks, and thus to ensure the
screening of any colour fields, it suffices to have the average
magnitude of the coupling in the InfraRed region as large as
$$
\lrang{\frac{\as}{\pi}}_{\mbox{\IR}}>
\frac{\al_{\mbox{\scriptsize crit}}}{\pi}\>=\> 
C_F^{-1}\left(1-\sqrt{
2/3}\right) \>\simeq\> 0.137\>.
$$
The interesting thing about this number
(apart from being easy to memorise) is that it is rather small.
\paragraph{Pragmatically,} 
if essential non-Abelian fields indeed grew really big, we would not
even know how to properly define gluon degrees of freedom because of
the notorious problem of ``Gribov copies''.\cite{copies}
In Hamiltonian language, the Coulomb QCD interaction in the presence
of transverse vacuum fields is described by the operator
\begin{equation}
  \label{eq:coulomb}
D^{-1}=\partial^2 \eta 
+g_s[{A_\mu^\perp},\partial_\mu\eta]\>, 
\end{equation}
reminiscent of the propagator of the Fadeev-Popov ghost in covariant
gauges.  In the second order in $g_s$, the statistical equal-time average
of the product of two vacuum fields, $\lrang{A_\mu^\perp
  A_\nu^\perp}\propto 4N_c$, takes over from the normal (unitary
respecting) screening effect due to the splitting of the Coulomb gluon
into ``physical fields'', namely two transverse gluons ($-N_c/3$) or a
$q\bar{q}$ pair ($-2n_f/3$).  

This physical explanation of the anti-screening
phenomenon,\cite{copies,Khriplovich}
had a dramatic continuation.  The normal magnitude
of field fluctuations of spatial size $L$ is $LA^\perp(L)\sim 1$. In
the background of very large vacuum fields, $LA^\perp(L)\sim g_s^{-1}$
(so large that the non-linearity becomes essential at the {\em
  classical}\/ level), the Coulomb (ghost) propagator
\eqref{eq:coulomb} becomes singular. Technically, this shows that we
did not manage to properly define Lagrangian physical degrees of
freedom, to divide out the volume of non-Abelian group
transformations, to fix the gauge.

A chilling perspective of Fadeev-Popov ghosts rising from the dead
makes one wish to get away with a numerically small coupling
(relatively small fields) as a unique (and maybe only) chance to keep
things under control.
\paragraph{Phenomenologically,} 
there is no sign of strong colour fields: confinement appears to be
{\em soft}\/ and {\em friendly}.
There are three aspects to this friendliness:
\begin{enumerate}
\item 
pQCD works OK from very large scales down to {$Q\!\!\sim$~2~GeV}, for the
phenomena where it 
{\em should}\/ (hadronic $\tau$-decays being an extreme example);
\item 
pQCD works down to (and below) {$Q\sim$ 1--2 GeV} where it 
{\em did not have to} (as it does, in particular, for DIS structure
functions at HERA);
\item 
  moreover, sometimes pQCD surprisingly works down to ...
  $k\!\!=\!\!\mbox{\scriptsize NILL}$. This happens, notably, in
  describing {\em inclusive}\/ energy and angular distributions of
  hadrons produced in hard processes, the phenomenon known as LPHD
  (local parton-hadron duality).\cite{gluer,KO}
\end{enumerate}
What are the main lessons we have learned from experiment? 
\begin{enumerate}
\item We have got used to pQCD covering orders of magnitude in the basic
  hard cross sections. More than that, our toddler-wisdom of how to
  estimate \NP\  
  effects is being certified. 
\item 
HERA tells us that proton is truly fragile. 
The transition from the $\gamma p$ physics to deep inelastic 
$\gamma^{*} p$ phenomena happens {\em early},  and it is {\em
  sharp}. The proton isn't actually bound, if you take my meaning: 
 a little scratch --- and it is blown to pieces.
\item Finally, when viewed {\em globally}, confinement is about {\em
    renaming}\/ a flying-away quark into a flying-away pion rather
  than about forces {\em pulling}\/ quarks back together.  From the
  study of numerous string/drag effects in particle flows, in $\epem$
  and at the Tevatron, we have learnt that even junky 200-300 MeV pions
  obediently follow the pattern of underlying colour fields.  Whatever
  the ultimate solution of the confinement problem may be, it had
  better be gentle in transforming the quark-gluon Pointing-vector
  into the Pointing-vector of the final state hadrons.
\end{enumerate}

\subsection{Some like it perturbative}
So, how to measure what we can't define?
The idea is to look for {\em deviations}\/ of 
inclusive quantities that characterise hard processes, from 
their respective
perturbative predictions. 
For a \PT-{\em calculable}\/
observable such a deviation,  
due to genuine \NP\ (confinement) physics,
is expected to be inverse proportional to
a certain {\em power}\/ of the hardness scale, 
$$
\frac{\delta \sigma^{\NP}}{\sigma} \>\propto
 \frac{\log^q Q}{Q^{2p}} \>.
$$
From within the \PT-approach these (observable-dependent) powers can
be inferred. 
Already at this stage we get a lot of valuable information.  For
example, 
\begin{itemize}
\item 
we would have no chance to see one of the most precise
determinations of $\amz$ from the $\tau$-decay if confinement effects
(which {\em apriori}\/ could have been huge at as small a scale as
$m_\tau<$~2~GeV) were not strongly suppressed~\cite{taupower,BBB} as
$(\LQCD^2/m_\tau^2)^3\ll1$; 
\item 
an understanding of the leading \NP\ 
power correction to the Drell--Yan $K$-factor,\cite{BB,DYK} $1/Q^2$,
calls for proper attention being paid to formulating the \PT\ 
predictions for production of Drell--Yan lepton pairs, heavy flavours
and high-invariant-mass jet pairs in hadronic
collisions~\cite{CMNThadro} in order to avoid an artificial $1/Q$
contamination;\cite{Berger}
\item 
discriminating jet-shape variables according to the envisaged power of
the leading \NP\ 
contribution allows the selection of observables less affected by
hadronisation and thus better suited for precise QCD tests such as
measuring $\as$.
\end{itemize}
As we shall discuss below, jet shapes
typically contain $1/Q$ \NP\ 
contributions.  Two examples of ``cleaner'' jet observables: the mean
value of the three-jet resolution variable $y_3$, defined according to
the Durham algorithm, which is hadronisation-sensitive at the $\log
Q/Q^2$ level,\cite{WDM} or {\em central moments},\cite{central}
$\lrang{(V-\lrang{V})^n}$, of most jet-shape observables 
$V$.\footnote{but not the broadenings, see later}
 
However the story does not end here.
Not only can one trigger the exponent of the confinement contribution to
a given observable but, 
making further assumptions,  one may quantify the 
{\em absolute magnitude}\/ of the \NP\ 
effects by relating the deviations found for different observables.

The programme can be looked upon as {\em pushing forward the
  Sterman-Weinberg wisdom}. \PT-calculable observables are 
Collinear-and-InfraRed-Safe (CIS) observables, those which can be
calculated in terms of quarks and gluons without encountering either
collinear (zero-mass quark, gluon) or soft (gluon) divergences.
The procedure is straightforward. 
\begin{enumerate}
\item Choose a CIS observable and get hold of the corresponding 
      state-of-the-art \PT-prediction;\footnote{These days, two-loop plus
      all-log resummed with special care being taken of the
      coupling running with the internal gluon momentum scale.}
\item enjoy the beauty of the latter;
\item observe that it has no sense (with $\as(k^2)$ running through
  the ``Landau pole'');
\item force it to have one (kindly ask $\as(k^2)$ to behave);
\item as far as the \PT\ 
  prediction produces now a definite answer, given an \IR-finite
  $\as$, see what you have done: quantify the ignorance about $\as$ in
  the low momentum range;
\item do it again for other observables;
\item verify that your ignorance is {\em universal}\/ i.e.\ 
  observable-independent;
\item eat your hat if it isn't and switch to another business.
\end{enumerate}
In recent years this programme has been carried out for a large set
of practically interesting quantities including
non-singlet DIS structure functions,\cite{WDM,A2,DISnonsing} 
$\epem$~\cite{hadro,eefragm} and DIS fragmentation 
functions,\cite{DSW}
various jet shape characteristics 
(means and distributions in thrust, jet masses, $C$-parameter, jet
broadening) in $\epem$
annihilation~\cite{KS,eeshapes,NS,DokWeb97,Milan}
and DIS,\cite{WebberDIS,DISshapes}
to heavy hadron spectra from heavy-quark-initiated jets~\cite{HQloss} and 
angular jet profiles.\cite{jetprofile}
First encouraging steps have been made towards revealing the \PT-\NP\ 
interplay in hard small-$x$ phenomena such as high energy
$\gamma^*\gamma^*$ scattering cross section,\cite{Haut} and non-singlet
DIS structure functions.\cite{DISsing}

Among interesting confinement-sick quantities still queueing for
treatment are back-to-back energy-energy correlation and transverse
momentum spectra of Drell--Yan pairs (which should exhibit weird
non-integer $Q$-exponents), out-of-event-plane characteristics of two-
and three-jet events (such as $T_{minor}$ or oblateness), accompanying
$E_T$ hadron flows in DIS and hadronic collisions, and many others.

\section{Assessments give rise to Calculations}

How to quantify \NP\ 
corrections to hard observables?  We shall sketch solutions to five
major problems that arise on the way, namely
\begin{enumerate}
\item How to disentangle power corrections coming from \UV\ 
and \IR\ 
regions?
\item How to split the magnitude of the power term into an 
  observable-dependent \PT-{\em calculable}\/ factor and
  a {\em universal}\/ \NP\ 
parameter?
\item Is the latter really {\em universal}? 
  What is the {\em accuracy}\/ of the universality statement? 
\item How to merge \PT\ 
and \NP\ 
contributions to the full answer?
\end{enumerate}
These being understood, the time comes to worry about your hat...

\subsection{Problem \#\ 1: \IR\ vs \UV}
The trick of introducing small gluon mass $m$ into Feynman diagrams
can be used to probe contributions of small momentum
scales.\cite{hadro,BBZ,BB95b,BB,BBB}  Operationally, the procedure is quite
simple.  Consider the \PT\ 
correction to a given observable $\sigma$, given by one-loop Feynman
diagrams with one additional gluon, real or virtual. Real and virtual
contributions, $\sigma_r$ and $\sigma_v$, are \IR-sensitive.
Introducing a finite mass into the Feynman propagator of the gluon
regularises collinear and infrared divergences producing
$$
 \sigma_r \>,\>\> \sigma_v \>=\> \cO{\ln^2\epsilon} 
                            +\cO{\ln \epsilon} + \cO{1}\>, \quad
\epsilon\equiv \frac{m^2}{Q^2}\,.
$$ 
However, having chosen a CIS observable we have secured the cancellation of
divergent terms in the physical answer, so that
$$
\sigma= \sigma_r \>+\> \sigma_v \quad=\quad  \cO{1}\,,
$$ 
thanks to the Bloch-Nordsieck theorem. 
By examining, how fast does the $m$-dependent contribution to $\sigma$ 
vanish in the $\epsilon\to0$ limit, we can quantify sensitivity of a
given observable to large distances. To this end we have to look for  
contributions {\em non-analytic}\/ in $\eps$ 
in the origin,\cite{BBZ,BB95b,BB,BBB} 
while analytic corrections proportional to
integer powers of $m^2/Q^2$ come from the \UV\ 
momentum region, $k\sim Q$, in the Feynman integrals involving the
modified gluon propagator,
\begin{equation}
G(k^2)=\frac1{-k^2-i0} \>\Longrightarrow\>  \frac1{m^2-k^2-i0}\,.
\end{equation}
For example, the one-loop total $\epem$ annihilation cross section
into hadrons has the following structure, 
$$
\sigma = \sigma_{Born}\left(1 + \frac{C_F\as}{2\pi}\cR(\eps)\right),
\quad \cR(0)=\frac32\>.
$$
with $\cR(\eps)$ the ``characteristic function'' for the $\sigma_{e^+e^-}$ 
depending on the gluon mass.
Setting $\eps=0$ we recover the well-know first order \PT\ 
correction $\sigma_1/\sigma_{Born}=\as/\pi$.  The function $\cR$ is
known, and so is its small-$\eps$
behaviour,  
$$
 \cR(\eps) = \frac32 -\frac32 \eps^2 + \frac{11}9\eps^3
-\frac{2\eps^3}3\ln\eps  
\>+\>\cO{\eps^4\ln\eps}.
$$ 
This shows that all the $\cO{\eps}$ terms present in $\sigma_r$ and
$\sigma_v$ have cancelled in the sum, as well as the $\ln^2\eps$- and
$\ln\eps$-enhanced $\cO{\eps^2}$ corrections:
non-analyticity starts at the $\cO{\eps^3}$ level only.
This property can be easily inferred from a mysterious reciprocity
relation~\cite{WDM} whose origin remains to be understood,
$$ 
\cR(\eps)-\frac32 = \eps^2\left(\cR(\eps^{-1})-\frac32\right).
$$
The message that such a powerful extension of the Bloch-Nordsieck
wisdom~\cite{BB,BBZ} sends us is that the first genuine large-distance
contribution to $\sigma_{e^+e^-}$ appears as a $Q^{-6}$
power correction, corresponding to the $\lrang{(\bar\psi\psi)^2}$
vacuum condensate of dimension six, in the standard OPE
language.\cite{OPE,ITEP}  

We shall return to the issue of {\em non-analyticity}\/ in $m^2$ a
little later when we discuss the \NP\ 
trigger.  But first, a confession is due: the last thing one would
like to do is to make the gluon field massive. Violating sacred
non-Abelian gauge invariance eventually results in breaking the \UV\ 
renormalisability of the theory.

\paragraph{Analytic coupling and ``dispersive mass''.}
Fortunately the situation is not so scary.
What we actually need is not a real Lagrangian mass. It suffices to
take into consideration the fact that the gluon, as any other quantum
field, lives part-time in virtual states consisting of various parton
configurations ($q\bar{q}$, $gg$, etc).  These virtual transitions
appear as a renormalisation of the gluon field,
\begin{equation}
 G(k^2)=\frac1{-k^2-i0} \>\Longrightarrow\>  \frac{Z(k^2)}{-k^2-i0}\>, 
\end{equation}
which can be characterised in terms of the distribution over 
``virtual mass'' $m^2$ (gluon spectral function) via a dispersion relation.
It is this mass that makes the gluon ``heavy''. 

The factor $Z$ has a simple physical meaning: it can be identified
with the running coupling,
\begin{equation}
  \label{eq:Zal}
  \frac{Z(-k^2)}{Z(-\mu_R^2)} \>=\> \frac{\al(k^2)}{\al(\mu_R^2)}\>,
\end{equation}
with $\mu_R$ an arbitrary \UV-renormalisation point. 
Such an identification is straightforward in the Abelian theory: in
QED, due to the Ward identity, radiative corrections to
the wave function of a charged particle (electron) and to the vertex 
cancel out ($Z_1=Z_2$); what remains is the photon propagator, 
$Z\equiv Z_3$. 
In QCD the gluon which mediates interaction between ``charges'' is a
charged (coloured) particle itself. Therefore its wave function
renormalisation, unlike that of the photon, is gauge-dependent and has 
no direct meaning. What then enters into the definition of
the physical coupling is the gauge-invariant combination of the gluon
propagator and specific non-Abelian corrections to the interaction vertices,
which are independent of the nature (colour representation) of the
external charges,
\begin{equation}
  Z\>=\> \Gamma^{(\mbox{\scriptsize NA})}\cdot Z_3
         \cdot\Gamma^{(\mbox{\scriptsize NA})} \>.
\end{equation}
Modulo this subtlety, in QCD
we can still say that 
exchanging a gluon with 4-momentum $k$  brings in the propagator  
\begin{equation}
\label{eq:Gal}
  G(k) \>=\> \frac {\as(-k^2)}{-k^2-i0}
\end{equation}
(where we have dropped an irrelevant overall constant). 
Now comes the crucial assumption: we want (\ref{eq:Gal}) to make sense 
in the entire complex $k^2$-plane. We know sufficiently well how
$\as(-k^2)$ behaves in the deep Euclidean region, at large negative
$k^2$; we know next to nothing about small-$k^2$ region. However,
whatever the function $\as$ is, it had better respect
causality. Therefore we suppress the formal \PT\ 
tachion (Landau singularity) and choose the ``physical cut'' alone,
$0<k^2<\infty$, as a support for the dispersive relation,
\begin{equation}
  \label{eq:disp}
\as(q^2)= \int_0^\infty \frac{dm^2\>q^2}{(m^2+q^2)^2}\,\ae(m^2)\>.  
\end{equation}
Here $\ae$ is the dispersive companion of the standard
coupling, 
\begin{equation}
  \label{eq:aedef}
 \frac{d}{d\ln \mu^2} \ae(\mu^2) \>=\>
-\frac{1}{2\pi i}\, \mbox{Disc} \left\{\as(-\mu^2)\right\} .
\end{equation}
The dispersion relation (\ref{eq:disp}) can be formally inverted 
as the operator relation
\begin{equation}
  \label{eq:dispinv}
 \ae(\mu^2)= \frac{\sin(\pi\cP)}{\pi\cP}\, \as(\mu^2)\>, \quad 
 \cP=\mu^2\frac{ d}{d\mu^2}\>.  
\end{equation}
It shows that in the \PT\ 
region $\ae$ and $\as$ are pretty close:
$\ae(k^2)=\as(k^2)(1+\cO{\as^2})$.  One can design model expressions
for $\as$ and the corresponding $\ae$.  What matters in our discussion
is that the latter gives us a direct hold on the large-distance
contribution to the observable under study.  We are ready for a ``heavy
gluon''.  Substituting \eqref{eq:disp} into the gluon propagator
\eqref{eq:Gal} we can write
$$ 
 \frac {\as(-k^2)}{-k^2-i0} = 
\int_0^\infty \frac{dm^2}{m^2} \,\ae(m^2) \cdot \frac{-d}{d\ln m^2}
\frac{1}{m^2-k^2-i0} \>.
$$
Feynman diagrams contain an integration over the internal gluon
momentum $d^4k$.  We conclude that the answer can be represented as an
$\ae(m^2)$-weighted integral over the ``mass'' $m^2$ of the ($\ln
m^2$--derivative of) the \PT\ 
answer calculated by usual Feynman diagram techniques but with an {\em as
  if massive}\/ gluon.  Such a calculation produces an $m^2$-dependent
{\em characteristic function}\/ $\cF_V$ specific to a given observable
$V$.  In the first order in the running coupling $\cF$ is a function
of $\eps=m^2/Q^2$ and is given by the one-loop diagrams with a gluon
of mass equal to the dispersive variable, $0\le m^2 \le \infty$.  We
get
\begin{equation}
  \label{eq:aeint}
V(Q^2\!\!,x)=\!\! \int_0^\infty \!\!\frac{dm^2}{m^2} \ae(m^2)
\cFd_V(\eps,x);
\>\> 
\cFd \equiv \frac{-d\cF}{d\ln\eps},
\end{equation}
with $x$ a set of relevant dimensionless variables.
At this point \eqref{eq:aeint} is no different from the usual \PT\ 
answer.  The CIS nature of $V$ guarantees convergence of the $m^2$
integration: $\cFd$ vanishes as a power of $\eps$ ($\eps^{-1}$) in the
$\eps\to0$ ($\eps\to\infty$) limit.  Therefore the distribution
$\cFd_V(\eps)$ has a maximum at $\eps=C_V(x)=\cO{1}$, and the integral
is dominated by the large-momentum region $m^2\sim Q^2$.
Approximating $\ae(m^2)\simeq\as(Q^2)$ we easily reproduce the one-loop \PT\ 
answer,
\begin{equation}
  \label{eq:oneloopPT}
 \sigma_V(Q^2,x)\>\simeq\> \as(Q^2)\cF_V(0,x)\>.
\end{equation}
Using the observable-dependent position of the maximum of the
$m^2$-distribution as the scale for the coupling in
\eqref{eq:oneloopPT}, $\as(C_V(x)Q^2)$, does a better job since it
minimises higher order effects. In this respect the dispersive
technology is closely related to the idea of ``commensurate
scales''.\cite{commensurate}

\paragraph{Renormalons.}
We could have tried better still by taking into account \PT\ 
corrections coming from logarithmic running of the coupling around the
maximum. However there is an imminent danger around the corner.  The
deficiency of the \PT\ 
series becomes apparent from the fact that the high-order expansion
coefficients $R_k$ exhibit factorial growth.\cite{renormalons} This
is associated with both the \UV\ 
and the \IR\ 
integration regions in $\eps$.

At the one-loop level we can equate $\ae$ with $\as$ and substitute
for $\ae(m^2)$ the standard geometric series 
$$ 
\ae(m^2) \simeq \as \sum_{k=0}^\infty 
\left( \frac{\cb_0 \as}{4\pi} \;\ln\frac{Q^2}{m^2}\right)^k \!\!\!,
\quad \as\equiv \as(Q^2)\,.
$$ 
This would give 
$$ 
 V(Q^2,x)-\as(Q^2)\cF_V(0,x)\>\simeq\> \as
 \sum_{k=1}^\infty 
\left(\frac{\cb_0\as}{4\pi}\right)^{k}
\!\!R_k\,,
$$ 
with the expansion coefficients given by 
\begin{equation}
  \label{eq:ckdef}
R_k = \int_0^\infty \frac{d\eps}{\eps} \left(\ln\frac{1}{\eps}\right)^k
\cFd_V(\eps)\,.  
\end{equation}
The ultraviolet contribution to $R_k$ is estimated by integrating 
\eqref{eq:ckdef} over $\eps > 1$. If $\cFd(\eps)$ vanishes as
$\eps^{-q}$ at large $\eps$, one finds for large $k$
\begin{equation}
 R_k^{\UV}
 \sim \int_1^\infty \frac{d\eps}{\eps} 
\left(\ln\frac{1}{\eps}\right)^k
\;\eps^{-q} \> \sim\>(-q)^{-k}\,k!\,.  
\end{equation}
This corresponds to an {\em ultraviolet renormalon}.
Such an alternating series can be evaluated by Borel summation.
This is because in the \UV\ 
integration region the replacement $\ae(m^2) \to \as(m^2)$ is a
reliable approximation and the contribution of this region can in fact
be evaluated explicitly (e.g.\ numerically) without any expansion.

The infrared contribution to $R_k$ is estimated by integrating 
over the region $\eps<1$. 
The small-$\eps$ behaviour of the characteristic function can be
cast as
\begin{equation}
  \cFd_V(\eps) \>\simeq\> \eps^p\>f_V(\ln\eps)\,,
\end{equation}
with $f$ a polynomial at most quadratic in $\ln\eps$. Both the leading 
power $p$ and $f$ depend on the observable under consideration.
Substituting into \eqref{eq:ckdef} one finds 
\begin{equation}
R_k^{\IR}
= \int_0^1 \frac{d\eps}{\eps} \left(\ln\frac{1}{\eps}\right)^k
\;\eps^{p}\,f(\ln \eps) \>\sim\>p^{-k}\, k!
\end{equation}
We again find a factorial behaviour known as an {\em infrared renormalon}. 
In this case however the coefficients are {\em non-alternating}\/ 
and therefore the series is not Borel-summable.
Attempts to ascribe meaning to such a series (which cannot even pretend 
at the status of asymptotic) 
give rise to unphysical {\em complex}\/ contributions at the level of 
$Q^{-2p}$  terms.  This is generally interpreted as an intrinsic uncertainty 
in the summation of the perturbative series.
This problem is of a physical nature and cannot be resolved by
formal mathematical manipulations alone.
It requires genuinely new physical input to obtain a sensible answer. 


\subsection{Problem \#\ 2: \NP\ trigger}
To extract the \NP\ 
large-distance correction to a given observable we split the coupling
into two components,
$$
\as(k^2) = \al^{\PT}(k^2) + \al^{\NP}(k^2)\>,
$$
and use the latter as a trigger.
It should be made clear that such a splitting is symbolic: it 
represents the coupling not in terms of two {\em functions}\/
but rather of two {\em procedures}.  
This ambiguity will be dealt with
later on. For the time being let us just sketch these two procedures.
\begin{itemize}
\item   
  Having met $\al^{\PT}(k^2)$ under the integral we are advised to
  calculate it {\em perturbatively}, that is in terms of (not too
  long) a \PT-series at the point $k^2\sim Q^2$ that our integral
  ``sits'' around. At the same time we are supposed not to worry about
  our \PT-coupling being potentially sick in the \IR\ 
  momentum region.
\item   
  On the contrary, integrals with $\al^{\NP}(k^2)$ are believed to be
  determined by that very same \IR\ 
  region, below some finite few-GeV scale.
\end{itemize}
Since the integral of the type
\begin{equation}
  \label{eq:mels}
  \int_0^\infty \frac{dk^2}{k^2}\, \as^{\NP}(k^2)\cdot k^{2p}\>=\>
\>\left(\mbox{few 100s MeV}\right)^{2p}
\end{equation}
converges, it provides us with a {\em dimensionful}\/  constant 
(analogous to the ITEP vacuum condensates\cite{ITEP} for the OPE-controlled
Euclidean quantities). These are the \NP\ 
parameters that will determine the magnitude of $Q^{-2p}$-suppressed
contributions to CIS observables.

As long as $\as^{\NP}(k^2)$ falls fast enough, 
its dispersive companion   
$\ae^{\NP}(m^2)$, according to \eqref{eq:disp}, 
 should have {\em vanishing moments}
$$
\lim_{k\to\infty}k^{2p}\al^{\NP}(k^2)=0 
\> \Longleftrightarrow \>
\int_0^\infty \frac{dm^2}{m^2} m^{2p} \ae^{\NP}(m^2)=0 \,,
$$
at least for the first few integer values of $p$.  That is why substituting
$\ae^{\NP}$ for the full $\ae$ in \eqref{eq:aeint} singles out
specifically {\em non-analytic}\/ terms in the characteristic function
$\cFd_V$.  The leading non-analytic term in the $\eps\to0$ behaviour
of $\cFd_V$ determines then the power $p$ of the $1/Q^{2p}$ \NP\ 
contribution.

The usual view that the \NP\ 
component of the coupling decays fast at large momenta is rooted in
the standard ITEP-OPE ideology.\cite{ITEP} According to the ITEP
picture, \NP\ 
physics is related to large-wavelength smooth gluon and quark
fields.  The presence of such fields in the vacuum does not affect
the propagation of \PT\ 
quanta with large Euclidean momenta, and $\as(k^2)$ in the \UV\ 
domain in particular. If those fields were non-singular at short
distances, the quark and gluon propagators (and thus the running
coupling) would be subject to {\em exponentially small}\/ corrections
only.  Small-size instantons are believed to be the first (singular)
non-trivial vacuum fields that disturb the propagation of quarks and
gluons at the level of {\em power}\/ corrections $k^{-2p}$ with large
exponents $p>p_{\max}\sim\beta_0\sim9$.  Phenomenological victories of
QCD sum rules\footnote{{\em They}\/ [QCD sum rules] {\em do not
    explain how the infrared soup is cooked, but taking this fact for
    granted, they skillfully utilise the recipe}.\cite{QCD20} } are
impressive.\cite{QCD20,Shifman} This does not necessarily mean however
that the true separation between the \UV\ 
and \IR\ 
physics is indeed as sharp and deep as it is implied by the ITEP
picture.  A viable alternative could be that the gluon Green function
at large momentum $k$, for example, bears the memory of the \IR\ 
domain already at the level of the relative $k^{-4}$ term (the tail of
the \IR-singular gluon polarisation operator).\cite{Gribov} Such a
heretic proposal does not necessarily undermine the notion of the
gluon condensate, $\lrang{\as G^2}$. On the contrary, it can make it
``calculable''.\cite{George98} In what follows we shall concentrate on
the practically most important {\em smallest}\/ powers $p=\half,1$ and
leave to future disputes the intriguing question of the {\em depth}\/
of the separation between the \UV\ and \IR\ phenomena.

The non-analyticity of $\cFd$, necessary to generate \NP\ 
power correction, is typically of two kinds.  In the first case we
have an integer power $\eps^p$ accompanied by logarithm(s) of $\eps$
which induces non-analyticity.  This is the case of DIS structure
functions, the Drell-Yan $K$-factor, the width of hadronic
$\tau$-lepton decay and the total $\epem$ annihilation cross section,
etc.  For example, for the valence DIS structure function one has
$$
\lim_{m\to0}\cFd_{\mbox{\scriptsize DIS}}
\>=\> a(x)\cdot \frac{m^2}{Q^2}\ln\frac{Q^2}{m^2} 
\quad +\ldots 
$$
Secondly, we may have a half-integer $p$. This is the case for many
so-called jet-shape observables that characterise, in a CIS manner,
the structure of final states produced in hard processes. Thrust,
invariant jet masses, $C$-parameter, jet broadening, energy-energy
correlation (as well as a list of others not yet being discussed in
the literature) belong to the $p=\half$ class. These quantities should
embody $1/Q$ power effects due to confinement physics:\cite{KS,eeshapes}
$$
\lim_{m\to0}\cFd_V\>=\> {a_V}\cdot 
\sqrt{\frac{m^2}{Q^2}}\quad + \ldots 
$$

\paragraph{Thrust.}
Historically the first among the jet shapes addressed in this context
has been the {\bf thrust}, which measures the ``pencilness'' of the final state
system,
$$
T = \max_{\vec{n}} \frac{\sum_i |\vec{n}\cdot\vec{p}_i|}
{\sum_i |\vec{p}_i|}\>.
$$
The direction $\vec{n}$ that maximises $T$ is called the thrust axis. 
Two back-to-back particles with 4-momenta $p$ and $\bar{p}$ (or
clusters of particles with {\em parallel}\/ momenta) produced in the
centre of mass of $\epem$ annihilation with cms energy $Q$
correspond to $T\!=\!1$.  Thrust deviates from unity for two reasons. One
is \PT\ 
gluon bremsstrahlung. $T$ is obviously a CIS observable since neither
collinear parton splitting nor infinitely soft gluon radiation affect
its value. Therefore the \PT-component of the mean thrust is
\begin{equation}
  \label{eq:PTthrust}
\lrang{1-T}^{(\PT)}\>=\> \cO{\as}\>.  
\end{equation}
Another reason is pure hadronisation physics.  pQCD radiation being
switched off, two outgoing quarks are believed to produce two
narrow jets of hadrons which are uniformly distributed in rapidity and
have limited transverse momenta with respect to the jet axis
(Field-Feynman hot-dog, or a string).  Within a simplified ``tube
model'' with an exponential inclusive distribution of hadrons
$$
 \frac{dN}{d\eta dk_\perp} = \mu^{-1}\>\vartheta(\eta_{m}-|\eta|)
                            \>e^{-k_\perp/\mu}\>, \quad 
       \mu=\lrang{k_\perp}\>,
$$
we have 
\begin{eqnarray*}
\sum_i |\vec{p}_i| &=& 
2\!\!\int \!d\eta\,  dk_\perp\frac{dN}{d\eta dk_\perp}
 \> k_\perp\cosh\eta = 2\mu \sinh\eta_{m} =Q\,; \\
\sum_i |p_{zi}| &=& 2\!\!\int\! d\eta\,  dk_\perp \frac{dN}{d\eta dk_\perp}
 \> k_\perp\sinh\eta = 2\mu(\cosh\eta_{m}-1)\,,  
\end{eqnarray*}
where, for the sake of simplicity, we have treated hadrons as massless.
Constructing the ratio we obtain
$$
 T = \frac{\cosh\eta_{m}-1}{\sinh\eta_{m}} = 1- \frac{2\mu}{Q} 
+ \cO{\frac{\mu^2}{Q^2}}, 
$$
so that the departure of thrust from unity occurs at the $1/Q$
level, with the characteristic hadronisation transverse momentum as a
relevant scale,
\begin{equation}
\label{eq:NPthrust}
  \lrang{1-T}^{(\NP)} \simeq \frac{2\lrang{k_\perp}}{Q}\>. 
\end{equation}
Introducing finite hadron mass(es) does not change the result. 
It is from the study of hadronisation models that 
the $1/Q$ effects first came into focus.\cite{hadro}

The \PT\ 
approach normally would not provide us with such a dimensionful
parameter: gluon transverse momenta are broadly (logarithmically)
distributed which results in the mean $\lrang{k_\perp}\propto \as Q$,
in accord with \eqref{eq:PTthrust}.

However now we have a \PT-handle on the large-distance physics, thanks
to the ``gluon-mass'' trigger.  Contribution to thrust from a single
gluon with momentum $k$ reads, in terms of light-cone (Sudakov)
variables, $k=\al p+\cb \bar{p} + \bk_\perp$,
$$
 \delta (1-T) \>=\> {\min\{\alpha,\beta \}}\>.
$$
Calculating the characteristic function for $\lrang{1-T}$ one obtains
\begin{eqnarray*}
\cF_T &\simeq&
\frac{C_F}{\pi}\int\frac{d\alpha}{\alpha}\frac{d\beta}{\beta} dk_\perp^2
\delta(\alpha\beta Q^2-k_\perp^2-m^2)\cdot 
{\min\{\alpha,\beta \}} \\
&=& \frac{2C_F}{\pi\,Q}\int_0^{Q^2} 
\frac{dk_\perp^2}{\sqrt{k_\perp^2+m^2}}\>.
\end{eqnarray*}
Hence,
$$
\cFd\equiv -m^2\frac{d\,\cF}{dm^2} \simeq \frac{2C_F}{\pi}\>
\frac{m}{Q}\>,
$$
which is precisely the $\sqrt{m^2}$ non-analyticity we were
expecting.  It is easy to see that the leading non-analyticity is due
to the radiation of {\em soft}\/ gluons at {\em large angles}, 
$k_0\sim  k_\perp\sim m \ll Q$.

Since soft gluon radiation has in fact a classical nature and its
pattern is simple, as it is universal, the relative magnitudes of the \NP\ 
contributions to CIS jet shapes appear to be simple as well. For
example, for the {\em mean}\/ jet shapes one obtains
\begin{equation}
  \label{eq:means}
 \lrang{V}^{\NP} = \frac{a_V}{Q}\cdot \frac{C_F}{2\pi}
\int_0^\infty \frac{dm^2}{m^2}\sqrt{m^2}\> \ae^{\NP}(m^2)\>,  
\end{equation}
where the coefficients $a_V$ are simple numbers having a clear
geometric origin.

\subsection{Problem \#\ 3: Universality}
The very concept of the \IR-finite coupling would make no sense
without universality, i.e.\ if we had to introduce for each observable
a private phenomenological parameter to fix the magnitude of the
confinement contribution. Therefore it is natural that the
universality concept has been intensively argued
for.\cite{eeshapes,HQloss,UNIVER}

It was soon recognised, however, that the technologies based on the
``massive gluon'' trigger, whether the renormalon-motivated approach
tracing the series of fermion bubbles or the dispersive approach, are
intrinsically ambiguous since there is no unique prescription for
including finite-$m^2$ effects into the definition of the shape
variable. For example, a perfectly legitimate definition of thrust for
a 3-parton system that consists of massless $q\bar{q}$ and a {\em
  massive}\/ gluon, suggested by Beneke and Braun, has produced a
result differing by a factor $\simeq 1.8$ from the ``conventional''
\eqref{eq:means}.\cite{BB} The ``naive estimate'' that emerges as
a result of substituting a ``massive gluon'' for the real final-state
system of massless partons is fine for triggering the power of the
\NP-contribution but fails to predict its magnitude, the latter
remaining {\em prescription-dependent}.

Moreover, as has been already mentioned above, the running $\as(k^2)$
in Minkowskian observables emerges only at the {\em inclusive}\/ level, as
a result of an integration over positive virtualities of the gluon
decaying into final-state offspring partons.  Nason and Seymour
rightfully questioned the application of the inclusive treatment of
gluon decays. They pointed out that jet shapes are not truly inclusive
observables, since kinematics of the offspring matters for the $V$ value.
Therefore the configuration of offspring partons in the gluon decay
may affect the value of the power term at next-to-leading level in
$\as$, which a priori is no longer a small parameter since the
characteristic momentum scale is low.

Both these problems called for analysis of the \NP-effects at the
two-loop level. Such analysis has been performed, and the result came
out unexpectedly simple. It was shown that there exists a definite
prescription for defining the ``naive estimate'' of the magnitude
of the power contribution, such that the two-loop effects of {\em
  non-inclusiveness}\/ of jet shapes reduce to a {\em universal},
observable-independent, renormalisation of the ``naive'' answer by the
factor~\cite{Milan}
$$
\cM \>\simeq\> 1.76 \>\> {(1.67)}\quad \mbox{for} \>\>
n_f=3\>\> {(0)} \,.
$$
This is true for the \NP-effects in the thrust, invariant jet mass,
$C$-parameter and broadening distributions. The same factor (known as
the {\em Milan factor}\/) also applies to the energy-energy
correlation measure away from the back-to-back region as well as to
the linear jet-shape observables in DIS.\cite{DISshapes}

The universality of the Milan factor has three ingredients.  Firstly,
it relies on the universality of soft radiation, the latter being
responsible for confinement effects.  Secondly, it obviously
incorporates the concept of the universal the QCD interaction
strength, all the way down to small momentum scales (which has been
processed through the dispersive machinery).  Finally, it stems from a
certain {\em geometric universality}\/ of the observables under
consideration, which includes their linearity, collinear finiteness 
(convergence of rapidity integrals) and Lorentz invariance
(independence of the gluon decay matrix element on the parent gluon
rapidity).\cite{Milan}  

Strictly speaking, the accuracy of the Milan factor isn't great: the
next loop would bring in the correction 
$$
\cM^{\mbox{true}} =\cM^{{\mbox{2-loop}}}  
\left( 1+\cO{\frac{\as}{\pi}}\right)
$$
with $\as$ entering at {\em small scale}.  Hence, a $\sim 20\%$
uncertainty in the value of the $\cM$ factor cannot be excluded. 
At the same time, the above ingredients of its {\em
  universality}\/  seem to be general enough as to
ensure its validity even beyond two loops.

\subsection{Problem \#\ 4: Merging}
We agreed to treat $\as^{\NP}$ as a {\em procedure}\/ rather than a
function, and thus representing the \NP-component of the answer in the
form \eqref{eq:means} is not the end of the story.  It is
unsatisfactory in two respects. Psychologically, it does not satisfy
our curiosity about interaction strength. More importantly, 
it remains symbolic since its \PT-counterpart is given by a
renormalon-sick series. We need to marry the \PT\ and \NP\ 
components into a reasonable answer.  To this end we first trade
$\ae$ back for the standard coupling\footnote{One can avoid
  introducing $\ae$ in the first place and instead directly exploit analytic
  properties of $\as(k^2)$.\cite{DSW}} by invoking the identity
following from \eqref{eq:disp},
$$
\int_0^\infty \frac{dm^2}{m^2}\, m\, \ae^{\NP}(m^2)
= {\frac4{\pi}}\cdot\int_0^\infty dk\>\al^{\NP}(k^2)\>.
$$
Then, we truncate the integration at some arbitrary finite value
$\mu_I$ which is large enough to neglect the \NP\ 
interaction, $\as^{\NP}(k)\simeq0$, $k>\mu_I$, and get rid of
$\as^{\NP}$ by substituting
$$ 
I=\int_0^{\muI} dk\>\al^{\NP}(k^2) 
= \int_0^{\muI} dk\>\as(k^2) - \int_0^{\muI} dk\>\al^{\PT}(k^2)\,.
$$
The first integral quantifies the average interaction 
strength in the \IR, via 
$$
 {\alpha}_0(\muI) \>\equiv\> 
\frac 1{\muI} \int_0^{\muI} dk\>\as(k^2).
$$
The second  contribution should be calculated
{\em perturbatively}, as a series in $\as$.  
The series for this subtraction term is factorially divergent in high
orders, as is the basic series for $\lrang{V}^{\PT}$. 
Introduction of the \IR\ 
matching scale $\mu_I$ makes the {\em full}\/ combined answer
renormalon-free though.  Finally,
\begin{equation}
  \label{eq:meanV}
 \lrang{V} = \lrang{V}^{\PT}(\as) + a_V\cdot\cP\>,
\end{equation}
where the universal \NP\ 
parameter $\cP\sim 1/Q$ equals, to the second order in $\as$,
\begin{eqnarray}  
\label{eq:cPdef}
\cP \>&=&\> \left. \frac{4C_F\cM}{\pi^2}\frac{\muI}{Q}\cdot\right\{
{\al}_0(\muI) \nonumber \\
&& \left. - \left[\,
    \as+\cb_0\frac{\as^2}{2\pi}
\left(\ln\frac{Q}{\muI}+1+{\frac{K}{\beta_0}}\right)+\ldots\right]
\right\}  
\end{eqnarray}
Here $K$ is a known number depending on the scheme chosen for the \PT\ 
expansion parameter $\as=\as(Q^2)$. A residual $\mu_I$-dependence, at
the level of $\cO{\as^3\mu_I/Q}$, is the price for having a
renormalon-free answer. In principle it can be reduced by continuing
the \PT-series, both for $\lrang{V}^{\PT}$ and the subtraction in
\eqref{eq:cPdef}.  The coefficients $a_V$ in \eqref{eq:meanV} for mean
thrust, $C$-parameter and invariant total and heavy-jet masses in
$\epem$ annihilation are
\begin{equation}
\label{eq:aV}
\begin{array}{|l|c||c||c|c|} \hline 
  V=   &   \>1-T\>  & \>\> C\>\>  & \> M_T^2\>  & \> M^2_H\>   \\ 
\hline
   a_V=   & 2 &  3\pi  & 2 & 1 \\
\hline  
\end{array}
\end{equation}
The same \NP-parameter \eqref{eq:meanV} that governs the leading
confinement effect in the {\em means}\/ affects the jet-shape {\em
  distributions}\/ as well.  In the ``soft'' kinematical
region\footnote{at the expense of introducing an additional phenomenological
  function, the {\em shape function}, it is possible to extend the
  range down to $V\!=\!0$, as 
recently demonstrated by
G. Korchemsky in the test thrust case.\cite{Gregshape}}
$\LQCD/Q\ll V \ll V_{\max}$ it reveals itself as a {\em shift}\/ in
the pure perturbative spectrum,\cite{DokWeb97,KS}
\begin{equation}
  \label{eq:shift}
  \frac{d\sigma}{dV}(V)\>=\> \frac{d\sigma^{(\PT)}}{dV}(V-a_V\cP)\>.
\end{equation}
The phenomenology of $1/Q$ effects in jet-shape means and
distributions suggests
$$
\cA_1 \equiv \frac{C_F}{2\pi} \int_0^\infty \frac{dk^2}{k^2}\cdot
k\> \as^{\NP}(k^2) \>\simeq\> 0.2\mbox{--}0.25\, \mbox{GeV}\,.
$$
Leading power corrections to other observables are determined by
higher moments of $\al^{\NP}$.  In particular, studies of $1/Q^2$
power corrections to DIS structure functions allow the quantification of the
{\em second}\/ moment of the \NP-coupling, corresponding to $p=1$ in
\eqref{eq:mels},\cite{A2,DSW}
$$
 \cA_2 \equiv \frac{C_F}{2\pi} \int_0^\infty \frac{dk^2}{k^2}\cdot
k^2\> \as^{\NP}(k^2)
\>\sim\> 0.2\, \mbox{GeV}^2.
$$

\noindent
Dedicated discussions of different aspects of what we know, think we
know and think about $\as$ can be found in the literature. Among the
most recent are: OPE, duality and the ``physical
coupling'',\cite{dedicated} the simplest model for the \IR-finite
$\as$,\cite{dedicatedSS} a more sophisticated model which fits two
known moments, $p\!=\!\half$ and $p\!=\!1$, and respects
OPE,\cite{dedicatedW} discussion of the causality
issue,\cite{dedicatedGGK} instructive lessons about OPE, \IR-finite
coupling and the Landau pole from the $O(N)$
$\sigma$-model.\cite{dedicatedBBK}

\section{Calculations give rise to Comparisons}

The phenomenology of power-suppressed contributions to jet shapes had a
vibrant but somewhat troubled childhood.  Only thrust and
$C$-parameter remained unaffected by theoretical misconceptions (some
of which we are going to sort out below).

A ball-park value of $\alpha_0\simeq 0.5$ was repeatedly
emerging from the analyses of jet shapes in $\epem$ and DIS current
jets in the Breit frame.\cite{shapes97,aleph97,H197,H198,JADEalpha} A
typical resume of 1997 would run like ``{\em The concept of a
  `universal' Power Correction parameter $\alpha_0$ in DIS $ep$
  scattering and $\epem$ annihilation is supported}''.\cite{H197}
However the same H1 group was the first to complain about the expected
$\ln Q$ enhancement of the $1/Q$ correction to broadening~\cite{Milan}
which was inconsistent with experiment.
The puzzle was recently put under scrutiny and clarified by the
resurrected JADE collaboration.
A detailed analysis presented at the Montpellier QCD conference by
Pedro Movilla Fernandez showed that hadronisation effects in
broadening not only {\em shift}\/ the distribution to larger $B$
values (as it is the case for $1\!-\!T$ and $C$) but also {\em
  squeeze}\/ it.  As a consequence, fitting the data with a $\log
Q$-enhanced $B$-independent shift yields inconsistent results for the
total and wide-jet broadening distributions (marked ``old $B_T$'' and
``old $B_W$'' in Fig.~5). 
They are also far away, in the parameter
space, from the {\em consistent}\/ pair of the thrust and
$C$-parameter distributions (solid $T$ and $C$ ellipses in
Fig.~5).\cite{JADEmontp} Alarming experimental conclusions ``{\em
  inconsistent results for jet broadening variables}\/'' and ``{\em
  effective coupling moments incompatible with each other}\/'' (read:
{\em there is no universality}\/) made theoreticians jump on the train
and revisit the problem from their side.

\begin{figure}[ht]
\begin{center}
\epsfig{figure=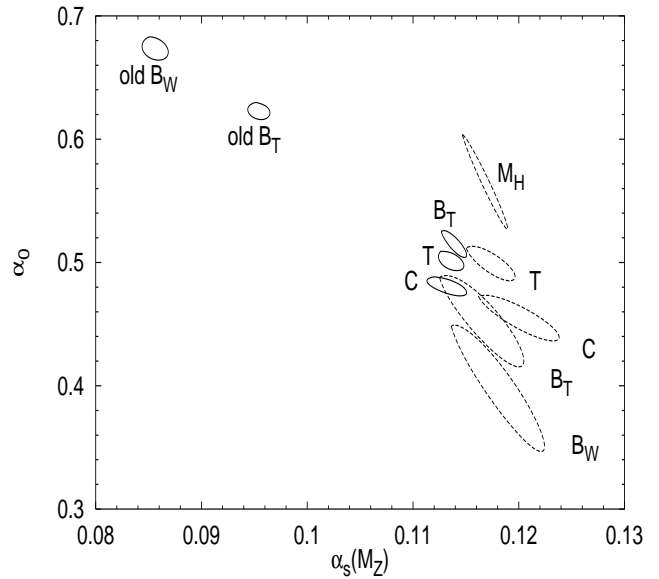,width=3.3in,height=3in}
\end{center}
\caption{95\% CL contours for jet shape
  means (dashed) and some distributions (solid).}
\end{figure}

It was soon recognised that one essential phenomenon was overlooked in the
original \NP-treatment of broadening,\cite{Milan} namely an
interplay between \NP- and \PT-phenomena.  To put it short, the effects
produced by \NP-radiation {\em in the presence of normal
  \PT-radiation}\/ are different from  the effects
of \NP-radiation inferred from a pure first-order analysis, that is
when the \PT-radiation is ``switched off''. 

\paragraph{\NP-effects in the presence of \PT-radiation.}
It is not for the first time that such ``mistake'' has been made. There
is a whole list of confusions that came from the first-order analysis
disregarding normal ever-present \PT-gluons. 

The story of the heavy-jet mass is the simplest example.  As you may
remember from the previous discussion, to trigger the \NP-contribution
we are advised to add to the parton system a soft {\em gluer},\cite{gluer}
a gluon with $k_\perp\sim m\sim \LQCD$.  Let us do so at the Born level,
that is add a gluer to the $q\bar{q}$ system as the third and only
secondary parton. Constructing $M^2$ of the quark-gluer system we will
find a $1/Q$ confinement contribution to the squared mass of the {\em
  heavy}\/ jet, the one our gluer belongs to.  Meanwhile, the opposite
{\em lighter}\/ jet containing a lonely quark gets none of it.  As a
result the \NP-correction to $M^2_H$ came out equal to that for
thrust, 
\begin{equation}
  \label{eq:wrongMH}
  a_T=a_{M^2_T}=a_{M^2_H}\,,\quad  a_{M^2_L}=0\>. \qquad\qquad \mbox{(wrong)}  
\end{equation}
In reality there are always normal \PT\ 
gluons in the game which are responsible for the
bulk of the jet mass: $M^2_H/Q^2\sim \as > M^2_L/Q^2\sim \as^2 \gg
\delta M^2_{\NP}$.  In these circumstances it is not gluer's business
to decide which of the jets is going to be heavier.
Confinement effects are instead shared equally,  see \eqref{eq:aV},
\begin{equation}
  \label{eq:rightMH}
  a_T=a_{M^2_T}=2a_{M^2_H}= 2a_{M^2_L}\>. \qquad\qquad \mbox{(right)}  
\end{equation}
It is worthwhile noticing that experimental analyses carried out before 1998
were based on the wrong expectation \eqref{eq:wrongMH}, which is
obviously not the experimenters' fault.\footnote{To the best of my
  knowledge the latest JADE analysis is the first one that properly
  included the Milan factor and fixed the $M_H$ confusion.\cite{JADEalpha}}

Another still popular ``mistake'' is to expect that the
\NP-contributions to jet shapes can be suppressed by measuring 
{\em higher moments},  for example,
$\lrang{(1\!-\!T)^n}$. 
The one-gluon analysis indeed would formally produce for such an observable
$$
\lrang{(1\!-\!T)^n}_{\mbox{\NP\ only}} \>\>\simeq\> \frac{\cA_n}{Q^n}.
$$
However what we are dealing with in reality instead is
$$
(1\!-\!T)^n = \left[\,(1\!-\!T)_{\PT} +
    (1\!-\!T)_{\NP}\,\right]^n .
$$
Symbolically,
$$
\lrang{\left((1\!-\!T)_{\PT} +
    \frac{\LQCD}{Q}\right)^{\!\!n} } \simeq \as + \as\frac{\LQCD}{Q} +
  \ldots  \left(\frac{\LQCD}{Q}\right)^{\!\!n} \!\!\!.
$$
The leading $1/Q$ contribution is still here, reduced by the $\as(Q^2)$
factor but far more important than the $1/Q^n$ term.

Another ``mistake'' of this sort
brings us closer to the $B$-issue.  Consider the transverse momentum
broadening of the current-fragmentation jet in DIS, that is the sum of
moduli of transverse momenta of particles in the current jet.  Adding
a gluer to the Born (parton model) quark scattering picture we get
{\em three}\/ equal contributions to $B$: two contributions from the
quark $p$ which recoils against the gluer $k$ emitted either in the
initial (IS) or in the final (FS) state,
$|\vec{p}_\perp|=|\vec{k}_\perp|$, and one contribution from the gluer
itself when it belongs to FS.  Taking into account the \PT-radiation,
however, the FS quark has already got a non-zero transverse momentum,
$\p_\perp^{\PT}\sim \as\cdot Q$, a substantial amount compared to
$k_\perp\sim\LQCD$.  In this environment the direct gluer's
contribution is the only one to survive: the \NP-recoil upon the quark
gets degraded down to the $1/Q^2$ effect after the azimuthal average
is performed,
$$
\lrang{|\vec{p}_\perp|} = \lrang{|\vec{p}_\perp^{\>\PT}-\vec{k}_\perp|}
= p_\perp^{\PT} +\cO{\frac{\LQCD^2}{p_\perp^{\PT}}}.
$$
The true magnitude of the $1/Q$ contribution turns out to be a factor 
{\bf three} smaller than that extracted from the one-gluon analysis.

Now we are ready to address the {\em squeezed broadening}\/
issue.\cite{Brevisit}

The feature that $1\!-\!T$ and $C$ have in common is that the dominant
\NP-contribution to these and similar shapes is determined by
radiation of gluers at {\em large}\/ angles.  This radiation is
insensitive to the tiny mismatch, $\Theta_q=\cO{\as}$, between the quark
and thrust axis directions which is due to \PT\ 
gluon radiation.  Therefore the quark momentum direction can be
identified with the thrust axis.

The broadening, on the contrary, accumulates contributions which do
not depend on rapidity, so that the mismatch between the quark and the
thrust axis starts to matter both in the $B$-means and distributions.
 
If one naively assumes that the quark direction coincides with that of
the thrust axis, then $B$ accumulates \NP-contributions from gluers
$i$ with rapidities up to the kinematically allowed value $\eta_i\le
\eta_{\max}\simeq\ln(Q/k_{ti})$.  In this case one finds the shift in
the $B$-spectrum to be loga\-rithmically enhanced,
\begin{equation}
  \label{old}
\Delta_B \>=\> a_B\cP \cdot \ln \frac{Q}{Q_B}\>,
\end{equation}
where $a_B=1(\half)$ for the total (single-jet; wide-jet)
broadening.\cite{Milan} What was overlooked here is the fact
that the {\em uniform}\/ distribution in $\eta_i$ (defined with
respect to the thrust axis) holds only for rapidities not
exceeding $|\ln \Theta_q|$.
{\em High-energy}\/ gluers with $k_{0i}>k_{ti}/\Theta_q$ are collinear 
to the {\em quark}\/ direction rather than to that of the thrust axis        
and therefore do not contribute essentially to $B$.
As a result, the \NP-contribution to $B$ comes out proportional
to the quark rapidity,
\begin{equation}
\label{eq:nptheta}
  \delta B_1^{(\NP)}   \>\simeq\> a_1\cP\cdot \ln\frac1{\Theta_q}\>.
\end{equation}
How does this affect \NP-contributions to $\lrang{B}$ and to the
$B$-distributions?  The power correction to the mean single jet
broadening $\lrang{B}_1$ is obtained by evaluating the perturbative
average of $\delta B_1$ in \eqref{eq:nptheta},
\begin{equation}
  \lrang{\delta B}_1^{(\NP)}  
  \>\simeq\> a_1\cP\cdot \lrang{\ln\frac1{\Theta_q}}.
\end{equation}
At the Born level, the \PT-distribution in the quark angle $\Theta_q$
is singular at $\Theta_q\!=\!0$. In high orders this singularity is
damped by the double-logarithmic Sudakov form factor.  As a result,
the \NP-component of $\lrang{B}_1$ gets enhanced by
\begin{equation}
   \lrang{\ln\frac1{\Theta_q}} \>\simeq\> \frac{\pi}{2\sqrt{C_F\as(Q)}}\>. 
\end{equation}
For the mean wide jet broadening $\lrang{B}_W$
the result has the same structure
with the replacement $C_F\to 2C_F$ due to the fact that now it is
radiation off {\em two}\/ jets which determines the $\Theta_q$ distribution.

The shift in the {\bf single jet} (wide jet) broadening can be
expressed as
\begin{equation}
  \Delta_1(B)   \>\simeq\> a_1\cP\cdot\lrang{\ln\frac1{\Theta_q}}_B\>,
\end{equation}
where the average is performed over the perturbative distribution in
the quark angle $\Theta_q$ while keeping the value of $B$ fixed.
Since $\Theta_q$ is kinematically proportional to $B$, the
$\log$-enhancement of the shift in the $B$-spectrum becomes
\begin{equation}
\label{new}
    \Delta_1(B) \>\simeq\> a_1\cP\cdot\ln\frac{B_0}{B}
\end{equation}
(with $B_0$ a calculable function slowly dependent on $\as\ln B$.
Thus, the shift in the $B_1$ ($B_W$) distribution becomes
logarithmically dependent on $B$.

The shift in the {\bf total} two-jet broadening distribution
$\Delta_T(B)$ has a somewhat more complicated $B$-dependence.  In the
kinematical region where the multiplicity of gluon radiation is small,
$\as\ln^2B\ll1$, one of the two jets is responsible for the whole
$\PT$-component of the event broadening, while the second is
``empty''. That ``empty'' jet contributes the most to the shift: in
the absence of perturbative radiation the direction of the quark
momentum in this jet stays closer to the thrust axis.  This results in
$$ 
  \Delta_T(B) \>\simeq\> \Delta_1(B) + \lrang{B}_1^{(\NP)} \>\simeq\>
  a_1\cP\left( \ln \frac1{B} +  \frac{\pi}{2\sqrt{C_F\as}} \right).
$$ 
In these circumstances the $B$-dependence of the total shift practically
coincides with that of a single jet.
In the opposite regime of well developed \PT-radiation, 
$\as\ln^2B\gg1$, 
the jets are forced to share $B$ equally, and we have instead
$$ 
   \Delta_T(B) \>\simeq\> 2\cdot \Delta_1(B/2) \simeq 
2\cdot a_1\cP\> \ln \frac1{B} \,.
$$ 
It is the $\log B$-enhancement of the \NP\ 
shift that makes the final distribution {\em narrower}\/ than
its \PT-counterpart. 

On July 27$^{\mbox{\scriptsize th}}$ when this talk was being
delivered, the origin of the ``mistake'' was already clear but the
true answer still unknown.  Therefore it wasn't risk-free to bet that
the $B$-distributions would eventually agree with respectable $T$ and
$C$.
Fig.~5, which is preliminary, shows $95\%$ CL contours for $\amz$ and
$\alpha_0$ as extracted from $B_T$ distributions in the energy range
35--183~GeV, and also from the mean values of $C$, $T$, $M_H^2$, $B_T$
and $B_W$. Some care is to be taken in interpreting these results,
since no account has been taken of the correlation of systematic
errors.\cite{Salam}

The universality of confinement effects has moved on to firmer ground, and
with it, the concept of an \IR-finite coupling.

\section{Comparisons give rise to Victories}

In our pursuit of confinement effects
we were following the logic of the great Chinese warrior-philosopher
Master Sun Tzu who said: 

\begin{flushright}
\begin{minipage}{3.0 in}
{\em The rules of the military are five:
  measurement, assessment, calculation, comparison and victory.  The
  ground gives rise to measurements, measurements give rise to
  assessments, assessments give rise to calculations, calculations give
  rise to comparisons, comparisons give rise to victories}.\cite{AW}
\end{minipage}
\end{flushright}

\noindent
In our context, it would be a bit premature to talk about victories yet.
It will suffice to have noticed that
\begin{itemize}
\item   
  QCD is alive and remains a rich source of technical and conceptual
  theoretical problems, of experimental challenges and phenomenological
  amusement;
\item 
things are orderly in the ``Hard Domain'', 
and $\as(Q^2)$ among them;
\item 
a crazy idea of probing {\em perturbatively}\/ 
the ``Soft Domain'' has gained ground;
\item   
  the pQCD-motivated technology for triggering and quantifying, in a
  universal way, genuine  confinement effects in hard observables is under
  construction,
\item 
  and that the effective large-distance interaction strength
  inferred from these studies,
  ${\lrang{\frac{\as}{\pi}}_{\mbox{\IR}} \simeq 0.14 - 0.17}$, turns out
  to be sufficiently {\em small}\/ as to apply \PT-language, at least
  semi-quantitatively, down to small momentum scales.
\end{itemize}

\begin{flushright}
\begin{minipage}{2.4 in}
{\em According to my assessment, even if you have many more troops than 
  others, how can that help you to victory?}\/~\cite{AW}
\end{minipage}
\end{flushright}

\noindent
Heat is building up, and QCD is about to undergo a {\bf faith
  transition}: we are getting ready to convince ourselves to talk
about ``{\em quarks and gluons}\/'' down to, and into, the InfraRed.  This
is the core of the Gribov programme of attacking the colour
confinement.\cite{Gribov} It is interesting to remark that the average 
\IR\ 
coupling, though numerically rather small, appears at the same time to
be just {\em large enough}\/ to activate the super-critical
light-quark confinement mechanism. Is it a mere coincidence?  Could
be. To answer the question, the last Gribov works should be understood
and developed (the write-up of the second paper concluding his
two-decade QCD study remained unfinished). This won't be easy, given
the complexity of the problem and, sadly, stone-solidness of our
prejudice.\footnote{It was solid enough already 20 years ago when the
  seminal paper on what is known now as Gribov copies, G.
  horizon,\cite{copies} (459 SPIRES citations, Dec.\ 98) was
  initially rejected by Nucl.Phys.B. So why wonder that his last
  paper,\cite{Gribov} submitted posthumously, was (politely) rejected
  just on the grounds that {\em it tries to address the confinement
    problem perturbatively}\/!}

\begin{flushright}
  {\em When on surrounded ground, plot. \\
    When on deadly ground, fight}.\cite{AW}
\end{flushright}
It is worth to keep in mind that on the QCD ground, as we know it
today, to {\em fight}\/ isn't enough. To {\em plot}\/ is often
necessary.  There is hardly a theorem around for which a proof could
be concluded by the respectable {\bf QED} = {\bf Q}uod {\bf E}rat {\bf
  D}emonstrandum (what had to be shown). The abbreviation {\bf QCD}
suits more, {\bf Q}uod {\bf C}onvenit {\bf D}emonstrandum (what was
{\em agreed}\/ to be shown).  A call for imagination and intuition,
and guts to defend them, is what singles out the QCD island of the SM
archipelago, the island where you never get bored.

\section*{Acknowledgements}
pQCD is a vast subject. My sincere apologies to those of my
colleagues, theoreticians and experimenters, whose essential
contributions to the field I may have overlooked or did not appreciate
enough because of lack of competence.  The right things displayed in
this review I have learned from Volodya Braun, Vladimir Gribov,
Georges Grunberg, Valery Khoze, Pino Marchesini, Al Mueller, Gavin
Salam, Bryan Webber, Kolia Uraltsev, Arkady Vainshtein and many
others.  The wrong things remain solely the author's responsibility.
Last but not least, I am deeply grateful to Siggi Bethke for
invaluable on-line help in a long course of preparing this talk
and to Gavin Salam and Emma Ciafaloni for their patience in teaching
English and Latin grammar, respectively.

\end{document}